\documentclass[journal,onecolumn]{IEEEtran}
\usepackage[caption=false, font=footnotesize]{subfig}
\usepackage{rotating}
\usepackage{pdflscape}
\usepackage{booktabs}
\usepackage{siunitx}
\usepackage{mathtools}
\usepackage{cuted}
\usepackage{multirow}
\usepackage{eqnarray}
\usepackage{amsmath}
\usepackage{placeins}
\usepackage{float}
\usepackage{lipsum}
\usepackage{flushend}
\usepackage{balance}
\usepackage{scalerel}
\usepackage{stfloats}
\usepackage{bigints}
\usepackage{textcomp}
\usepackage{tabularx}
\usepackage{cleveref}
\usepackage{booktabs}
\usepackage{graphicx}
\usepackage{makecell}
\usepackage{siunitx}
\usepackage{tcolorbox}
\usepackage{enumerate}
\usepackage{stackengine}

\usepackage[export]{adjustbox}

\usepackage{multicol}
\usepackage{nomencl}
\usepackage{etoolbox}
\makenomenclature
\renewcommand\nompreamble{\begin{multicols}{2}}
\renewcommand\nompostamble{\end{multicols}}

\usepackage{acronym}
\usepackage{array}

\begin{document}

\title{Hybrid FSO and RF Lunar Wireless \\ Power Transfer}

\author{Barış~Dönmez,~\IEEEmembership{Graduate Member,~IEEE,}
        Yanni~Jiwan-Mercier,~\IEEEmembership{Graduate Member,~IEEE,}
        Sébastien~Loranger,~\IEEEmembership{Senior~Member,~IEEE}
        and~Güneş~Karabulut~Kurt,~\IEEEmembership{Senior~Member,~IEEE}
\thanks{B. Dönmez, Y. Jiwan-Mercier, S. Loranger, and G. Karabulut Kurt were with the Poly-Grames Research Centre, Department of Electrical Engineering and Space Resources and Infrastructure Engineering Research Unit (ASTROLITH), Polytechnique Montréal, Montréal, QC.
Corresponding Author E-mail: baris.donmez@polymtl.ca}
}


\maketitle

\begin{abstract}
This study focuses on the feasibility analyses of the hybrid free-space optics (FSO) and radiofrequency (RF)-based wireless power transmission (WPT) system used in the realistic Cislunar environment, which is established by using system tool kit (STK) high-precision orbit propagator (HPOP) software in which many external forces are incorporated. In our proposed multi-hop scheme, a solar-powered satellite (SPS) beams the laser power to the low lunar orbit (LLO) satellite in the first hop, then the harvested power is used as a relay power for RF-based WPT to two critical lunar regions, which are lunar south pole (LSP) (0$^\circ$E, 90$^\circ$S) and Malapert Mountain (0$^\circ$E, 86$^\circ$S), owing to the multi-point coverage feature of RF systems. The end-to-end system is analyzed for two cases, \textit{i}) the perfect alignment, and \textit{ii}) the misalignment fading due to the random mechanical vibrations in the optical inter-satellite link. It is found that the harvested power is maximized when the distance between the SPS and LLO satellite is minimized and it is calculated as 331.94 kW, however, when the random misalignment fading is considered, the mean of the harvested power reduces to 309.49 kW for the same distance. In the next hop, the power harvested by the solar array on the LLO satellite is consumed entirely as the relay power. Identical parabolic antennas are considered during the RF-based WPT system between the LLO satellite and the LSP, which utilizes a full-tracking module, and between the LLO satellite and the Malapert Mountain region, which uses a half-tracking module that executes the tracking on the receiver dish only. In the perfectly aligned hybrid WPT system, 19.80 W and 573.7 mW of maximum harvested powers are yielded at the LSP and Mountain Malapert, respectively. On the other hand, when the misalignment fading in the end-to-end system is considered, the mean of the maximum harvested powers degrades to 18.41 W and 534.4 mW for the former and latter hybrid WPT links.
\end{abstract}


\section*{}
\label{Nomen}

\nomenclature{$I(r,z)$}{Laser irradiance as a function of $r$ and $z$} \nomenclature{$r$}{Radial distance from the centre of the beam} 
\nomenclature{$z$}{LoS distance between the SPS laser and LLO solar array centre}   
\nomenclature{$I_0$}{Maximum irradiance at the beam center} 
\nomenclature{$w_0$}{Beam waist, or $w(0)$} 
\nomenclature{$w(z)$}{The beam radius limited by $1/e^2$ at $z$} 
\nomenclature{$\eta_{eo}$}{PCE from electrical power to optical power} 
\nomenclature{$P_S$}{Total input (electrical) power of the SPS} 
\nomenclature{$\lambda_o$}{Optical wavelength} 
\nomenclature{$\theta$}{Beam divergence angle} 
\nomenclature{$\lambda_o$}{Optical wavelength}
\nomenclature{$D_o$}{Transmitting telescope lens diameter}
\nomenclature{$P_{R_l}$}{Received optical power in perfect alignment}
\nomenclature{$b$}{Radius of the solar array}
\nomenclature{$A$}{Area of the circular solar array}
\nomenclature{$v$}{Random radial misalignment in FSO WPT}
\nomenclature{$P_{M_l}$}{Received optical power in misalignment}
\nomenclature{$\sigma_o$}{Pointing error standard deviation in FSO WPT}
\nomenclature{$f_{v}(v)$}{PDF of $v$}
\nomenclature{$\beta_o$}{Pointing error angle in FSO WPT}
\nomenclature{$P_{H_l}$}{Harvested electrical power by the LLO satellite}
\nomenclature{$\eta_{oe}$}{PCE from optical power to electrical power}
\nomenclature{$P_T$}{Total input (electrical) power of LLO satellite}
\nomenclature{$P_{E_l}$}{Harvested electrical power by the LLO satellite with a pointing error}
\nomenclature{$\eta_{er}$}{PCE from electrical power to RF signal power}
\nomenclature{$P_R$}{Received RF power}
\nomenclature{$d$}{LoS distance between the parabolic antennas located on the LLO satellite and the Moon}   
\nomenclature{$\lambda_r$}{RF signal wavelength}
\nomenclature{$G_R$}{Receiver dish antenna gain}
\nomenclature{$G_T$}{Transmitter dish antenna gain}
\nomenclature{$G_a$}{Parabolic antenna gain}
\nomenclature{$\rho_a$}{Parabolic antenna efficiency}
\nomenclature{$\phi_a$}{Angle off boresight}
\nomenclature{$D_a$}{Parabolic antenna diameter}
\nomenclature{$J_1(\cdot)$}{First-order Bessel function}
\nomenclature{$P_{H_p}$}{Harvested DC power by LSP antenna}
\nomenclature{$P_{H_m}$}{Harvested DC power by Mons Malapert antenna}
\nomenclature{$\eta_{re}$}{PCE from RF signal power to electrical power}
\nomenclature{$P_{R_p}$}{Received RF power by LSP antenna}
\nomenclature{$P_{R_m}$}{Received RF power by Malapert Mountain antenna}
\nomenclature{$P_{M_p}$}{Received RF power by LSP antenna in laser pointing error case}
\nomenclature{$P_{M_m}$}{Received RF power by Malapert Mountain antenna in laser pointing error case}
\nomenclature{$P_{E_p}$}{Harvested DC power by LSP antenna in laser pointing error case}
\nomenclature{$P_{E_m}$}{Harvested DC power by Mons Malapert antenna in laser pointing error case}
\nomenclature{$d_p$}{LoS distance between the parabolic antennas located on the LLO satellite and at LSP} 
\nomenclature{$d_m$}{LoS distance between the parabolic antennas located on the LLO satellite and at Mons Malapert} 
\nomenclature{$f_r$}{RF signal frequency} 
\nomenclature{$D_R$}{Receiver dish antenna diameter}
\nomenclature{$D_T$}{Transmitter dish antenna diameter}
\nomenclature{$\rho_R$}{Receiver parabolic antenna efficiency}
\nomenclature{$\rho_T$}{Transmitter parabolic antenna efficiency}
\nomenclature{$\phi_T$}{Transmitter angle off boresight}
\nomenclature{$\phi_R$}{Receiver angle off boresight}
\nomenclature{$G_{T_m}$}{LLO satellite-Malapert Mountain link transmitter antenna gain}
\nomenclature{$G_{R_m}$}{LLO satellite-Malapert Mountain link receiver antenna gain}
\nomenclature{$G_{T_p}$}{LLO satellite-LSP link transmitter antenna gain}
\nomenclature{$G_{R_p}$}{LLO satellite-LSP link receiver antenna gain}

\printnomenclature
\section*{Abbreviation}
\label{Accron}
\begin{acronym}
\acro{AoI}\,\,\,\,\,\,\,\,\,\,{Angle of Incidence} 
\acro{DC}\,\,\,\,\,\,\,\,\,\,{Direct Current}
\acro{DRO}\,\,\,\,\,\,\,\,\,\,{Distant Retrograde Orbit}
\acro{ELO}\,\,\,\,\,\,\,\,\,\,{Elliptical Lunar Orbit}
\acro{EMLP}\,\,\,\,\,\,\,\,\,\,{Earth-Moon Lagrange Point}
\acro{FLO}\,\,\,\,\,\,\,\,\,\,{Frozen Lunar Orbit}
\acro{FSO}\,\,\,\,\,\,\,\,\,\,{Free-Space Optics}
\acro{HPOP}\,\,\,\,\,\,\,\,\,\,{High Precision Orbit Propagator}
\acro{LLO}\,\,\,\,\,\,\,\,\,\,{Low Lunar Orbit}
\acro{LoS}\,\,\,\,\,\,\,\,\,\,{Line-of-Sight}
\acro{LSP}\,\,\,\,\,\,\,\,\,\,{Lunar South Pole}
\acro{MVA}\,\,\,\,\,\,\,\,\,\,{Moon Village Association}
\acro{NRO}\,\,\,\,\,\,\,\,\,\,{Near Rectilinear Orbit}
\acro{PCE}\,\,\,\,\,\,\,\,\,\,{Power Conversion Efficiency}
\acro{PCO}\,\,\,\,\,\,\,\,\,\,{Prograde Circular Orbit} 
\acro{PDF}\,\,\,\,\,\,\,\,\,\,{Probability Density Function}
\acro{RF}\,\,\,\,\,\,\,\,\,\,{Radiofrequency}
\acro{SPS}\,\,\,\,\,\,\,\,\,\,{Solar-Powered Satellite} 
\acro{STK}\,\,\,\,\,\,\,\,\,\,{System Tool Kit}
\acro{SWIPT}\,\,\,\,\,\,\,\,\,\,{Simultaneous Wireless Information and Power Transfer}
\acro{WPT}\,\,\,\,\,\,\,\,\,\,{Wireless Power Transmission}
\end{acronym}

\IEEEpeerreviewmaketitle

\section{Introduction}

\IEEEPARstart{T}{he} Moon, which is the Earth's natural satellite, has a lot of promise for a variety of reasons. First of all, it permits expeditions beyond the Cislunar realm, like deep-space missions (i.e., interplanetary) because it can be used as a stopover point. It is possible to maintain a spaceship or a lunar rover. Moreover, the MVA aims to settle and explore the Moon through collaboration with hundreds of participants from many nations. Furthermore, the precious minerals on the crust of the Moon attract private enterprises for the lunar mining opportunities \cite{coggins2024nasa,donmez2024continuous,MoonVillageAsc,peacock2017mining}. 

There are multiple promising regions on the LSP and Mons Malapert is one of them. The summit of Malapert Mountain has an altitude of 4,700 m and is located at latitude 86$^\circ$S and longitude 0$^\circ$E which is inside of the LSP. This mountain offers many advantages, such that it enables a continuous LoS channel with the Earth and hence it can be considered as a relay base. Unlike the Shackleton crater, which is also one of the favourable LSP locations, Malapert has relatively flat and large lunar terrain that facilitates the approach and landing with radar systems. Moreover, lunar mining can be carried out easily due to the form of the crust covering its surface. Furthermore, there is a negligible amount of interference induced by Earth-based orbital communication \cite{lowman2008making,mazarico2011illumination}.

There are various Cislunar orbit types such as LLO, PCO, FLO, ELO, NRO, EMLP-2 Halo, and DRO. The perilune of the circular LLO is approximately 100 km with an orbit period of around two hours. Besides, LLO has no constraint on its inclination, however, it is one of the unstable lunar orbits \cite{whitley2016options}. Due to its lower altitude and any inclination possibility, LLO facilitates the lunar surface access as well as the signal transmission.

In WPT, there are multiple options, such as the ubiquitous RF and novel FSO. The latter technology becomes advantageous when the distances widen, as the detector is able to collect a relatively large portion of the collimated beam transmitted by the laser, whereas the random pointing error induced by the mechanical vibrations shall be considered, as these misalignment losses can be significant \cite{donmez2023mitigation}. On the other hand, RF transmission through longer distances (e.g., kilometres) degrades the received signal power fairly, however, it enables multiple access due to its larger beam footprint compared to FSO. 

\subsection{Related Works}
In the existing works, the FSO technology is preferred more frequently as the distances in the space environment are ultra-long. In \cite{williams1993power}, FSO-based WPTs from EMLP-1 and -2 to a manned lunar rover are studied to harvest 30 kW of power. In \cite{williams1990diode}, equally spaced three satellites orbiting around the Moon with an altitude of 2,000 km are studied for harvesting 75 kW and 1 MW of power by a rover and a lunar habitat, respectively. In \cite{kerslake2008lunar}, the power efficiencies of cable, RF, and FSO are compared for surface-to-surface WPT on the Moon. In \cite{donmez2023mitigation}, FSO-based inter-satellite WPT considering random misalignment losses are studied. The transmitted power is collected by receiver telescopes having diameters of 0.1 m and 0.2 m on the 1U and 12U small satellites, respectively, and then the received optical power is harvested by solar cells. In \cite{donmez2024continuous}, the continuous WTP coverage on LFS is studied by assessing the multi-satellite halo orbit configurations on EMLP-2 and the losses induced by stochastic misalignment are incorporated in the analyses. In \cite{naqbi2024impact}, the negative impacts of lunar dust on power beaming are explored under the perfect alignment scenario. It is demonstrated that the attenuation is significant in a ground-to-ground WPT scenario. In \cite{lopez2023lunar}, a custom circular orbit, FLO, and DRO are considered up to ten satellites for WPT to the Shackleton crater and the lunar equator. Power budget analyses are carried out for different AoI constraints, and perfect alignment is assumed for the FSO. In \cite{pan2022space}, inter-satellite RF-based SWIPT and FSO-based SWIPT are compared regarding their features and architectures. The PCEs between DC and RF outperform the PCEs between DC and laser, whereas path loss is significantly higher for the RF WPT systems. In addition, RF-based systems enable point-to-multi-point WPT due to their larger coverage area than FSO-based systems can provide. 

\subsection{Motivation and Contributions}
In our proposed hybrid WPT model, a laser-based inter-satellite WPT is realized from an SPS to an LLO satellite in the first stage. Then, the LLO satellite operates as a relay which uses the harvested power and establishes RF-based WPT to two different points which are LSP (0$^\circ$E, 90$^\circ$S) and Malapert Mountain (0$^\circ$E, 86$^\circ$S). There are identical parabolic receiver antennas at these points, and power conversion is executed using a corresponding PCE. 

Our contributions are as follows
\begin{itemize}
  \item A realistic space environment in which the gravitational forces of third celestial bodies, SRP, and Moon radiation pressures such as thermal and albedo effects are taken into consideration is designed by using STK HPOP \cite{stk2024}.
  \item The influential variables for $P_{H_l}$, $P_{H_p}$, and $P_{H_m}$  are investigated. It was found that distance is an influential variable for computing both $P_{H_l}$ and $P_{H_p}$, whereas $G_{T_m}$ is the dominant variable for the evaluation of $P_{H_m}$. Thus, extreme cases are considered when statistical distributions of $P_{E_l}$, $P_{E_p}$, and $P_{E_m}$ are determined as a result of the random pointing error. 
  \item In the first hop, FSO-based WPT is realized between the SPS and LLO satellite. When there is no pointing error, $P_{H_l}$ can reach up to 331.94 kW and drop down to 305.33 kW. However, when the Rayleigh distributed random pointing error with $\sigma_o$ = 0.5 m is considered, the mean values of $P_{E_l}$ are 309.46 kW and 281.90 kW for $z_{min}$  and $z_{min}$, respectively.
  \item In the second hop, RF-based WPT is realized between the LLO satellite and two separate locations, which are the LSP and Mons Malapert. When there is no misalignment error and hence $P_T$ = $P_{H_l}$, the maximum values of $P_{H_p}$ and $P_{H_m}$ are 19.80 W and 573.7 mW, however, when $P_T$ = $P_{E_l}$, the mean values of $P_{E_p}$ at $d_{p,{min}}$ and $d_{p,{max}}$ are 18.41 W and 807.1 mW, respectively, and the mean values of $P_{E_m}$ at $G_{T_{m,max}}$ and $G_{T_{m,min}}$ are 534.3 mW and 66.06 nW, respectively. Since our performance analyses focus on the end-to-end hybrid WPT system, random $P_{E_l}$ values are computed by considering the $P_{T}$ value at the specific sample times at which $P_{H_p}$ and $P_{H_m}$ take extreme values. For instance, $G_{T_m}$ is the minimum at 00:31:50, thus, the value of $P_{T}$ at that very second is considered. The simulation outcomes of $P_{H_p}$, $P_{H_m}$, $P_{E_p}$, and $P_{E_m}$ represent the end-to-end hybrid WPT system harvested powers.
\end{itemize}

\section{Material and Method}

\begin{figure*}[h]
\centering
\includegraphics[width=.95\columnwidth]{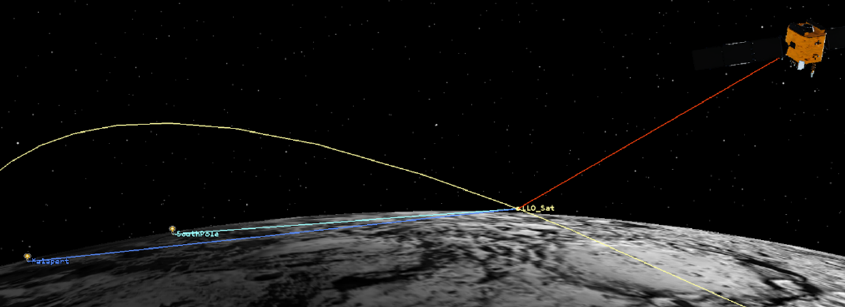}
\caption{In the first hop, SPS transfers laser power to the LLO satellite. This laser power is harvested by the solar array attached to the LLO satellite and is used as a relay power. In the second hop, RF-based WPT is realized between the LLO satellite and parabolic antennas located at LSP and Mons Malapert regions.}
\label{fig1}
\vspace{-0.2cm}
\end{figure*}

In our proposed hybrid WPT model shown in Fig. \ref{fig1}, FSO-based WPT between SPS and LLO satellite is realized with and without random misalignment error in the first hop. The harvested power by the LLO satellite is consumed as relay power for multi-point RF-based WPT in the last hop. As the path losses are significantly high through the lunar locations due to the extremely long distances (e.g., hundreds of kilometres), full-tracking ($\phi_{T}$ = $\phi_{R}$ = 0$^\circ$) is utilized for the LLO satellite and LSP link, which is the primary location, and semi-tracking ($\phi_{R}$ = 0$^\circ$) is used for the LLO satellite and Malapert Mountain as the transmitter cannot track two distant locations simultaneously. 

The Keplerian parameters used to establish the proposed system model are listed in Table \ref{table1}. The simulation duration is 2 hours, which is an approximate period of an LLO satellite, with a sampling time of 10 seconds. 

\newpage

\begin{table}[h]
\centering
\caption{Orbital Simulation Parameters}
\label{table1}
\begin{tabular}{lll}
\toprule
\toprule
Parameter                       & LLO satellite & SPS \\ 
\midrule
Semi-major axis                 & 1837.4 km      & 2037.4 km   \\
Eccentricity                    & 1.76383$\times$10$^{-32}$  & 2.15718$\times$10$^{-16}$   \\
Inclination                     & 90$^\circ$            & 90$^\circ$   \\
Argument of perigee             & 0$^\circ$             & 0$^\circ$   \\
Longitude of the ascending node & 355$^\circ$           & 90$^\circ$  \\
True anomaly                    & 180$^\circ$           & 180$^\circ$  \\
Satellite mass                  & 1000 kg        & 5000 kg  \\ 
\bottomrule
\bottomrule
\end{tabular}
\end{table}

The parameters of the FSO-based WPT simulation in which the random misalignment error is considered are presented in Table \ref{table2}. 

\begin{table}[h]
\centering
\caption{FSO-based WPT simulation parameters}
\label{table2}
\begin{tabular}{ccccccccc}
\toprule
\toprule
$P_{S}$  & $\lambda_{o}$  & $\eta_{eo}$  & $D_{o}$  & $\theta$  & $\eta_{oe}$  & $b$  & $\sigma_{o}$  & $\beta_{o}$  \\
\midrule
1 MW & 1064 nm & 51\% \cite{donmez2023mitigation} & 0.3 m & 3.547 $\mu$rad & 68.9\% \cite{lopez2023lunar} & 2 m & 0.5 m & 2.68 $\mu$rad  \\
\bottomrule
\bottomrule
\end{tabular}
\end{table}

The parameters of the RF-based WPT simulation in which multiple lunar regions are considered are presented in Table \ref{table3}.

\begin{table}[h]
\centering
\caption{RF-based WPT simulation parameters}
\label{table3}
\begin{tabular}{cccccccc}
\toprule
\toprule
$P_{T}$  & $f_{r}$  & $\eta_{er}$  & $D_{T}$  & $D_{R}$  & $\eta_{re}$  & $\rho_{T}$  & $\rho_{R}$  \\
\midrule
$P_{H_l}$ or $P_{E_l}$ & 2.5 GHz & 80\% \cite{pan2022space} & 4 m & 50 m & 80\% \cite{pan2022space} & 90\% & 90\% \\
\bottomrule
\bottomrule
\end{tabular}
\end{table}

There are multiple steps in our proposed hybrid FSO and RF WPT systems. The methods we use are explained briefly as follows
\begin{enumerate}[i]
  \item. The orbital simulation parameters shown in Table \ref{table1} are defined to be used in STK simulation in which many time-varying forces (e.g., third-body forces) are incorporated.
  \item. LoS distances (i.e., $z$, $d_{p}$, $d_{m}$), azimuth angles, and elevation angles (i.e., $\theta_{T}$, $\theta_{R}$) are obtained as a function of time by using STK.
  \item. The visibility intervals between the LLO satellite and SPS, LSP and Malapert Mountain are obtained from the STK simulation outcomes. Finding the common visibility interval shown in Fig. \ref{fig3} is necessary for comparing the RF-based WPT performances of LSP and Mons Malapert.
  \item. The parabolic antenna gains $G_{T_p}$, $G_{R_p}$, $G_{T_m}$, and $G_{R_m}$ are computed by using Eq. (\ref{eq10}) and Table \ref{table3}, then the outcomes are presented in Fig. \ref{fig4}.
  \item. The FSO-based WPT with perfect alignment at the LLO satellite is computed by using Eq. (\ref{eq4}) and (\ref{eq7}). Then the relationship between $P_{H_l}$ and $z$ is investigated as demonstrated in Fig. \ref{fig5}.
  \item. By using Eq. (\ref{eq9}--\ref{eq11}) with the consideration of $P_T$ = $P_{H_l}$, the RF-based WPT is realized for the regions mentioned in Step (iii). The harvested RF powers $P_{H_p}$ and $P_{H_m}$ and their relationships with the corresponding distances and antenna gains are investigated as shown in Fig. \ref{fig6}.
  \item. Using the Monte Carlo technique, many random misalignment errors $v$  are generated with the parameter $\sigma_{o}$ resulting from $\beta_{o}$  which are defined in Table \ref{table2}. The theoretical Rayleigh distribution curve validates the simulation results showing $f_{v}(v)$ in Fig. \ref{fig7} as well.
  \item. Section \ref{EH_Misalignment} explains the random misalignment fading with an illustration in Fig. \ref{fig2} and Eq. (\ref{eq5}--\ref{eq8}). The investigations in Steps (v) and (vi) to be able to find the influential variables of $P_{H_l}$, $P_{H_p}$ and $P_{H_m}$ are crucial as we consider two extreme cases which maximize and minimize $P_{E_l}$, $P_{E_p}$ and $P_{E_m}$. 
  \item. Once an inversely proportional relationship between $P_{H_l}$ and $z$ are concluded, the PDF of $P_{E_l}$ for the $z_{min}$ and $z_{max}$   extreme cases are presented in Fig. \ref{fig7}. Statistical values are extracted as well.
  \item. Then, an inversely proportional relationship between $P_{H_p}$ and $d_{p}$ is validated, too. However, $d_{m}$ is not the dominant factor for the $P_{H_m}$. A directly proportional relationship between $P_{H_m}$ and $G_{T_m}$ is found from Fig. \ref{fig6}. 
  \item. As we consider an end-to-end hybrid WPT system rather than two independent FSO-based and RF-based point-to-point systems, the sample times at which the maximum and minimum $P_{H_p}$ and $P_{H_m}$ occur are crucial since selecting the maximum $P_{T}$ by ignoring its time would be inconvenient. Thus, the sample times that are used during the computations of $P_{E_p}$ and $P_{E_m}$ in extreme cases must match with the sample time of the $P_{T}$ = $P_{E_l}$. The sample times at which extreme cases occur can be determined from Fig. \ref{fig6} and \ref{fig8}.
  \item. The statistical distributions of $P_{E_p}$ and $P_{E_m}$ for the two end-to-end hybrid WPT systems are presented in Fig. \ref{fig9} for the extreme cases. It should be noted that, as the RF frequency increases,  $P_{H_p}$ increases too, whereas $P_{H_m}$ decreases since the LLO satellite antenna, which tracks the dish at LSP, becomes more directive.
\end{enumerate}

 \section{Theory and Calculation}
 
\subsection{Laser Transmission Model}
The flux density, or the irradiance, of the laser can be modelled with a frequently considered Gaussian distribution. The Gaussian density profile decreases as the distance between the transmitter and receiver increases and can be defined as \cite{majumdar2010free}. 

\begin{equation}
I(r,z)={{I}_{0}}\left( \frac{w_{0}^{2}}{w{{(z)}^{2}}} \right)\exp \left( \frac{-2{{r}^{2}}}{w{{(z)}^{2}}} \right),
\label{eq1}
\end{equation}

\noindent where ${{I}_{0}}=\frac{2{{\eta }_{eo}}{{P}_{S}}}{\pi w_{0}^{2}}$. Hence, we can rewrite Eq. (\ref{eq1}) as follows

\begin{equation}
I(r,z)=\frac{2{{\eta }_{eo}}{{P}_{S}}}{\pi w{{(z)}^{2}}}\exp \left( \frac{-2{{r}^{2}}}{w{{(z)}^{2}}} \right),
\label{eq2}
\end{equation}

The beam (footprint) radius is determined from the point where the irradiance is reduced to 1/e$^2$ of ${{I}_{0}}$ and is defined by 

\begin{equation}
w(z)={{w}_{0}}\sqrt{1+{{\left( \frac{{{\lambda }_{o}}z}{\pi w_{0}^{2}} \right)}^{2}}},
\label{eq3}
\end{equation}

\noindent where ${{w}_{0}}=\frac{{{\lambda }_{o}}}{\pi \theta }$ and $\theta \cong {{\lambda }_{o}}/{{D}_{o}}$ \cite{ghassemlooy2019optical}.

\begin{figure*}[h]
\centering
\includegraphics[width=0.3\columnwidth]{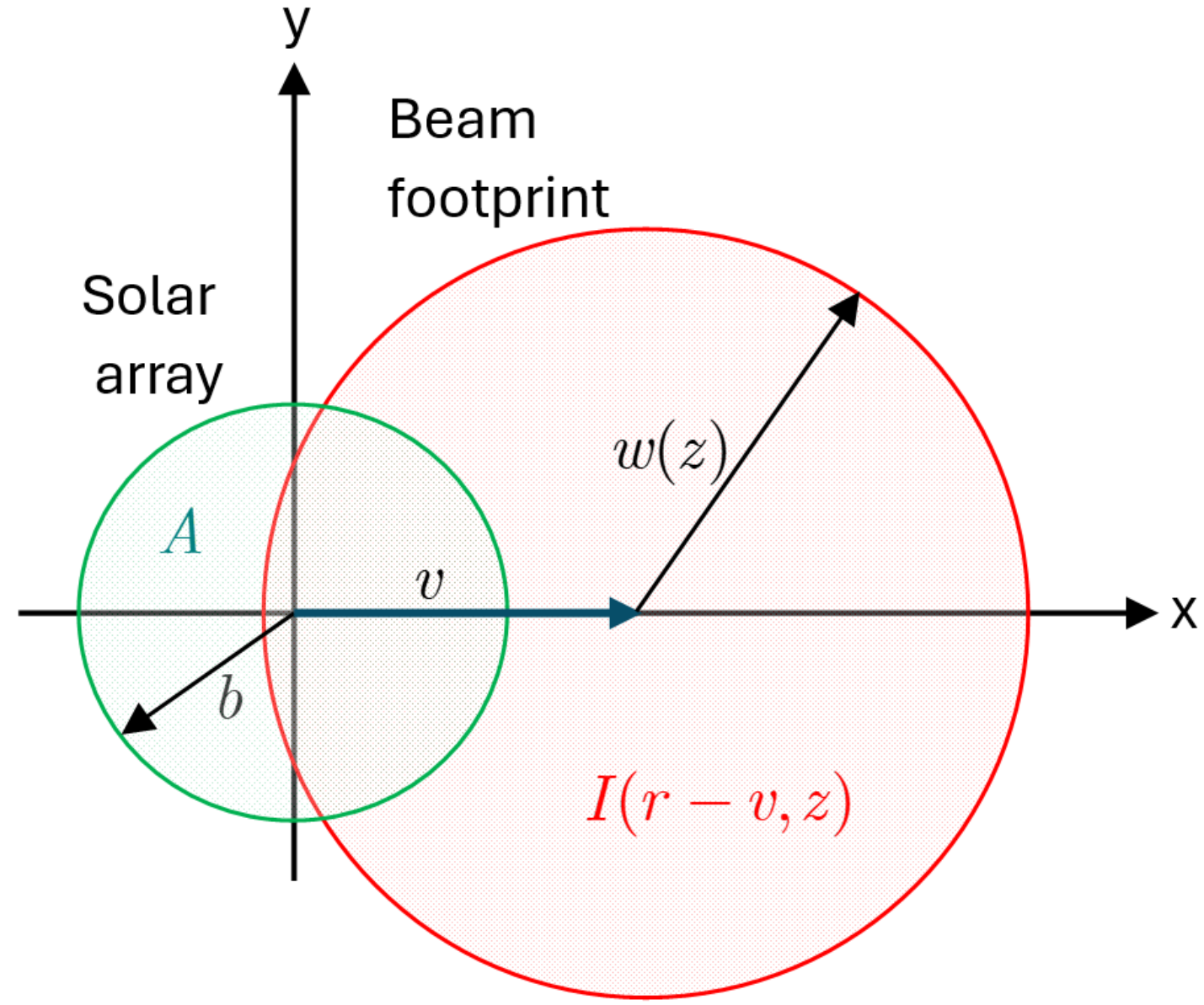}
\caption{Random pointing error $v$ shows the misalignment between the centres of the solar array with radius $b$ and randomly swayed irradiance $I(r-v,z)$ with a beam radius of $w(z)$.}
\label{fig2}
\vspace{-0.2cm}
\end{figure*}

\subsection{Laser Misalignment Model}
The misalignment of the beam footprint centre over the solar array/detector centre causes degradation in the collected portion of the transmitted power from the laser transmitter as the intersected region becomes smaller as presented in Fig. \ref{fig2}. Although the pointing error is a vector, the symmetry of the solar array's circular area and the circular shape of the beam footprint enables us to consider only its magnitude and hence the error vector is reduced to a scalar value. It can be visualized by rotating Fig. \ref{fig2} about the origin, and this will not change the collected amount of power. 

It can be assumed that the radial distance is placed on the x-axis. The received portion of the optical power when there is a perfect alignment is defined as follows  

\begin{equation}
{{P}_{{{R}_{l}}}}=\int\limits_{-b}^{b}{\int\limits_{-\sqrt{{{b}^{2}}-{{x}^{2}}}}^{\sqrt{{{b}^{2}}-{{x}^{2}}}}{\frac{2{{\eta }_{eo}}{{P}_{S}}}{\pi w{{(z)}^{2}}}}}\exp \left( -2\,\frac{\,{{x}^{2}}+{{y}^{2}}}{w{{(z)}^{2}}} \right)\,\,dy\,dx,
\label{eq4}
\end{equation}

The misalignment between the centres of the solar array and the beam footprint causes a loss, and Eq. (\ref{eq4}) is modified as follows \cite{farid2007outage}

\begin{equation}
{{P}_{{{M}_{l}}}}=\int\limits_{-b}^{b}{\int\limits_{-\sqrt{{{b}^{2}}-{{x}^{2}}}}^{\sqrt{{{b}^{2}}-{{x}^{2}}}}{\frac{2{{\eta }_{eo}}{{P}_{S}}}{\pi w{{(z)}^{2}}}}}\exp \left( -2\,\frac{{{(x-v)}^{2}}+{{y}^{2}}}{w{{(z)}^{2}}} \right)\,\,dy\,dx
\label{eq5}
\end{equation}

\noindent where $v$ can be modelled stochastically with Rayleigh distribution when the vertical and horizontal displacements on the receiver plane are considered as independent and identical Gaussian distributions \cite{farid2007outage} 

\begin{equation}
{{f}_{v}}(v)=\frac{v}{\sigma _{m}^{2}}\exp \left( -\frac{{{v}^{2}}}{2\sigma _{m}^{2}} \right),\,\,\,\,\,\,\,\,\,v>0
\label{eq6}
\end{equation}

\subsection{Harvesting the Optical Energy}
The harvested electrical power by the LLO satellite can be determined after the conversion of the received optical power to the electrical power as follows:

\begin{equation}
{{P}_{{{H}_{l}}}}={{\eta }_{oe}}\,{{P}_{{{R}_{l}}}}
\label{eq7}
\end{equation}

\begin{equation}
{{P}_{{{E}_{l}}}}={{\eta }_{oe}}\,{{P}_{{{M}_{l}}}}
\label{eq8}
\end{equation}

\subsection{RF Transmission Model}
The RF signal transmission is modelled with the Friis equation \cite{friis1946note} in which $P_{T}$ = $P_{H_l}$ or $P_{T}$ = $P_{E_l}$ as the total harvested power is used as the transmit power of the LLO satellite.

\begin{equation}
{{P}_{R}}=\left( {{\eta }_{er}}{{P}_{T}} \right)\,{{\left( \frac{{{\lambda }_{r}}}{4\pi d} \right)}^{2}}{{G}_{T}}{{G}_{R}}
\label{eq9}
\end{equation}

\noindent where $G_{T}$ or $G_{R}$ can be computed as a function of $\phi_{a}$ as follows \cite{gagliardi2012satellite} 

\begin{equation}
{{G}_{a}}({{\phi }_{a}})={{\rho }_{a}}{{\left( \pi \frac{{{D}_{a}}}{{{\lambda }_{r}}} \right)}^{2}}\left( \frac{2\,{{J}_{1}}(\zeta )}{\zeta } \right)
\label{eq10}
\end{equation}

\noindent where $\zeta =\frac{\pi {{D}_{a}}}{{{\lambda }_{r}}}\sin {{\phi }_{a}}$.

\subsection{Harvesting the RF Energy}
The harvested electrical power by an antenna located on the Moon can be determined after the conversion of the received RF power to the electrical DC power as follows:

\begin{equation}
{{P}_{{{H}_{p}}}}={{\eta }_{re}}\,{{P}_{{{R}_{p}}}}\text{\,\,\,\,\,\,\,\,\,\,or\,\,\,\,\,\,\,\,\,\,}{{P}_{{{H}_{m}}}}={{\eta }_{re}}\,{{P}_{{{R}_{m}}}}\text{ }
\label{eq11}
\end{equation}

\begin{equation}
{{P}_{{{E}_{p}}}}={{\eta }_{re}}\,{{P}_{{{M}_{p}}}}\text{\,\,\,\,\,\,\,\,\,\,or\,\,\,\,\,\,\,\,\,\,}{{P}_{{{E}_{m}}}}={{\eta }_{re}}\,{{P}_{{{M}_{m}}}}\text{ }
\label{eq12}
\end{equation}

\section{Results and Discussion}
\begin{figure*}[h]
\centering
\includegraphics[width=0.45\columnwidth]{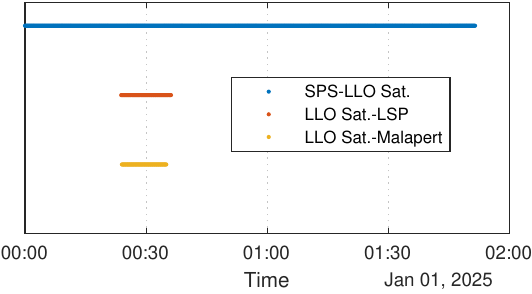}
\caption{All WPT link visibility intervals are used for finding a common time interval for the comparison of two end-to-end WPT at LSP and Mons Malapert.}
\label{fig3}
\vspace{-0.2cm}
\end{figure*}

The connections between SSP to LLO satellite, LLO satellite to LSP, and LLO satellite to Malapert Mountain have limited duration, and the corresponding visibility intervals are exhibited in Fig. \ref{fig3}. The visibility durations between LLO satellite to LSP and Mountain Malapert are 12:20 min. and 11:00 min., respectively. However, the common time interval for end-to-end WPTs is considered between 00:24:00 and 00:35:00, which lasts 11 minutes, for a fair comparison between LSP and Mons Malapert. 

\begin{figure*}[h]
\centering
\includegraphics[width=0.55\columnwidth]{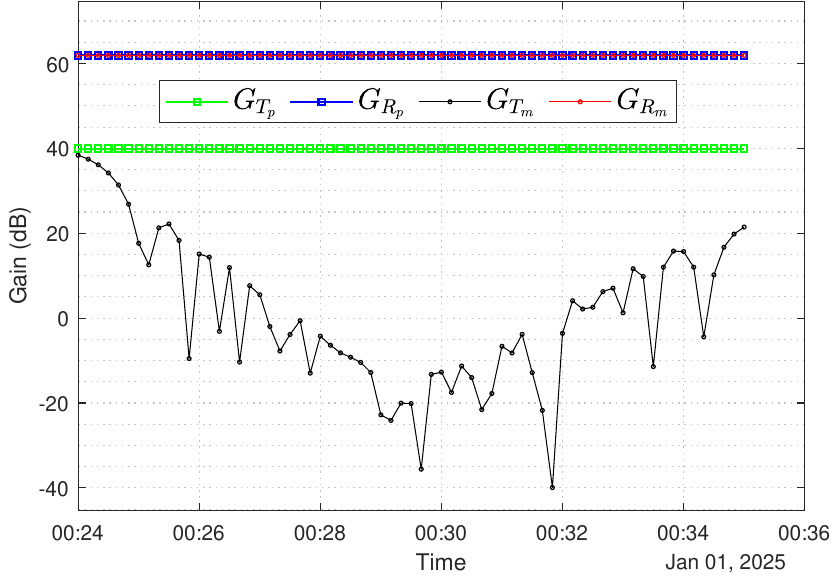}
\caption{Transmitting and receiving parabolic antenna gains as a function of time for the RF-based WPT realized at LSP and Malapert Mountain.}
\label{fig4}
\vspace{-0.2cm}
\end{figure*}

The parabolic antennas placed on the LLO satellite and the LSP track each other, and hence $G_{T_p}$ and $G_{R_p}$ can be maximized at 39.95 dB and 61.89 dB, respectively, as presented in Fig. \ref{fig4}. However, the Malapert Mountain, in which another dish antenna is located, is also able to collect RF signals from the same LLO satellite transmitter focusing on another point on the Moon, owing to the multi-point transmission feature of the RF systems. The $G_{T_m}$ varies as a function of $\phi_{T}$, between 38.37 dB and -39.92 dB whereas $G_{R_m}$ is maximized as 61.89 dB as presented in Fig. \ref{fig4} since both of the receiving lunar dish antennas use tracking devices for maximizing the RF-based energy harvesting from the relaying satellite. 

\subsection{Energy Harvesting with Perfect Alignment}
\label{EH_Perfect}
\begin{figure*}[h]
\centering
\includegraphics[width=0.32\columnwidth]{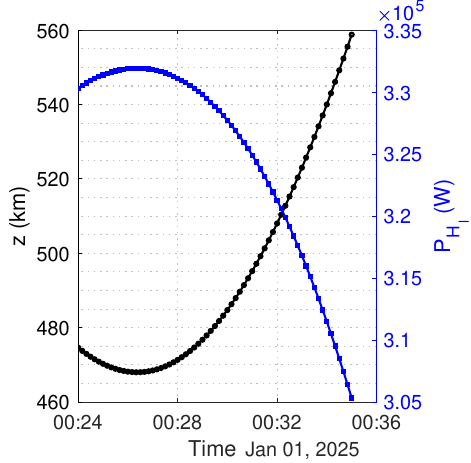}
\caption{The inversely proportional relationship between harvested power $P_{H_l}$ by the LLO satellite and SPS-LLO satellite LoS distance $z$ in FSO-based WPT.}
\label{fig5}
\vspace{-0.2cm}
\end{figure*}

\begin{figure*}[h]
\centering
\subfloat[]{
	\label{subfig:6a}
	\includegraphics[clip, scale=0.73]{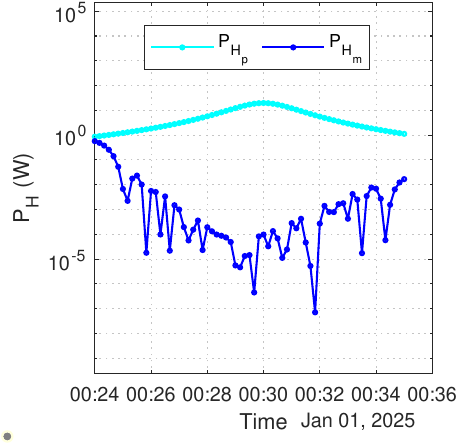}
	 }
\subfloat[]{
	\label{subfig:6b}
	\includegraphics[clip, scale=0.73]{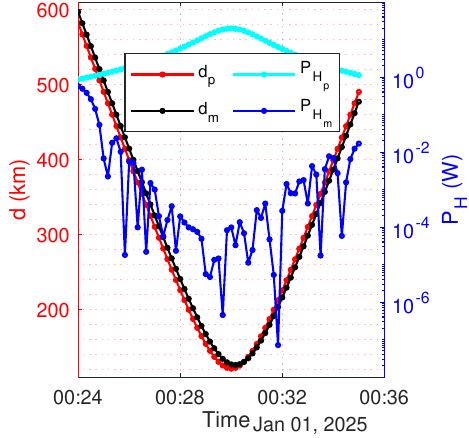}
	 }
\subfloat[]{
	\label{subfig:6c}
	\includegraphics[clip, scale=0.73]{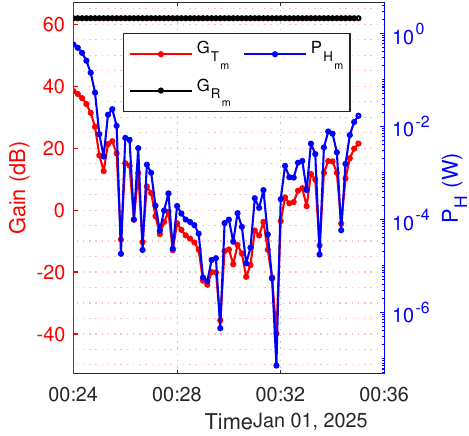}
	 }
\caption{(a) The comparison between $P_{H_p}$ and $P_{H_m}$, (b) the relationship between distances and harvested powers, and (c) the relationship between $P_{H_m}$ and antenna gains}
\label{fig6}
\vspace{-0.15cm}
\end{figure*}

The received power $P_{R_l}$ at the LLO satellite is then harvested by the circular solar array. $P_{H_l}$ is computed by using Eq. (\ref{eq4}) and is presented in Fig. \ref{fig5}. In addition, $P_{H_l}$ increases as $z$ decreases, thus, it is convenient to focus on the minimum and the maximum $z$ in the following subsection, yet the distance may not be the dominating variable for RF-based harvested powers as there are significant variations in the $G_{T_m}$. The minimum and maximum $P_{H_l}$ are 305.33 kW and 331.94 kW, which are necessary as harvested power will be used as relay power in our RF-based WPT and the path loss will be significant, especially for hundreds of kilometres of links. As $P_{T}$ = $P_{H_l}$, the harvested powers $P_{H_p}$ and $P_{H_m}$, which are computed using Eq. (\ref{eq9}--\ref{eq11}), are presented in Fig. \ref{fig6}.

\newpage
$P_{H_p}$ is inversely proportional to $d_{p}$ as exhibited in Fig. \ref{fig6}. For instance, 19.80 W is harvested when the distance is 121.34 km at the time 00:30, or $P_{H_p}$ is minimized at 00:24. However, this is not the case for $P_{H_m}$ and $d_{m}$, the maximum $P_{H_p}$ of 573.7 mW is obtained when $G_{T_m}$ is maximum, which is 38.37 dB (See Fig. \ref{fig4}), at time 00:24 although $d_{m}$ becomes maximum, which is 597 km. 

The average values of $P_{H_p}$ and $P_{H_m}$ are 6.05 W and 30.5 mW, and this discrepancy occurs because the LLO satellite antenna’s boresight only tracks the antenna's boresight at the LSP. 

\begin{figure*}[h]
\centering
\captionsetup{justification=centering}
\subfloat[]{
	\label{subfig:7a}
	\includegraphics[clip, scale=0.78]{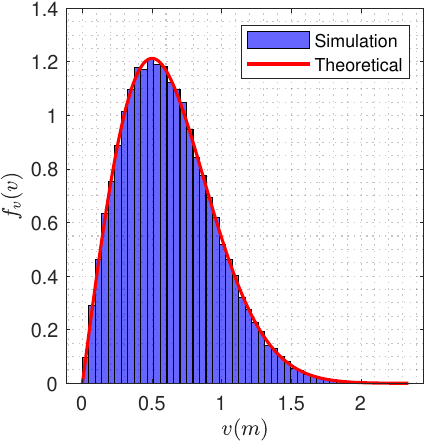}
	 }
\subfloat[]{
	\label{subfig:7b}
	\includegraphics[clip, scale=0.78]{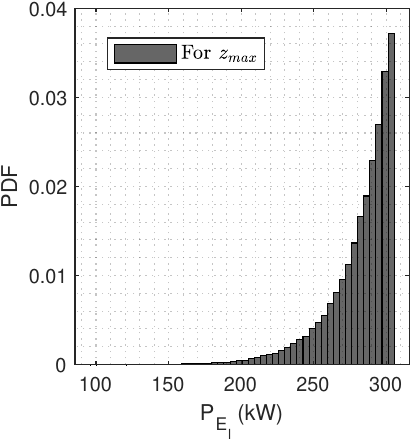}
	 }
\subfloat[]{
	\label{subfig:7c}
	\includegraphics[clip, scale=0.78]{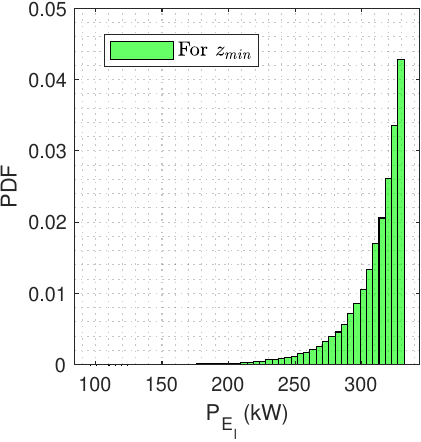}
	 }
\caption{Stochastic behaviours of (a) the pointing error $v$ on the circular solar array, (b) harvested power $P_{E_l}$ for the maximum $z$, and (c) harvested power $P_{E_l}$ for the minimum $z$.}
\label{fig7}
\vspace{-0.15cm}
\end{figure*}

\subsection{Energy Harvesting with Random Misalignment}
\label{EH_Misalignment}
Many random pointing errors with $\sigma_{o}$ = 50 cm are generated with the Monte Carlo approach first. Simulation results are validated by the theoretical Rayleigh distribution in Eq. (\ref{eq6}), and then this is demonstrated in Fig. \ref{subfig:7a}. Furthermore, the stochastic distributions of $P_{E_l}$ for the maximum and minimum $z$ are presented in Fig. \ref{subfig:7b}, and Fig. \ref{subfig:7c}, respectively. The mean values of $P_{E_l}$ for the $z_{max}$ and $z_{min}$ are 281.93 kW and 309.49 kW, respectively.

The harvested powers $P_{E_p}$ and $P_{E_m}$ will have statistical distributions as well due to random $P_{T}$ = $P_{E_l}$. In Section \ref{EH_Perfect}, the influential parameters which take $P_{H_p}$ and $P_{H_m}$ to the maximum and the minimum are investigated. According to Fig.~\ref{fig5}~and~\ref{fig6}, $P_{H_p}$ is inversely proportional to $d_{p}$ whereas $P_{H_m}$ is directly proportional to the dominant variable $G_{T_m}$. Hence, statistical distributions of $P_{E_p}$ and $P_{E_m}$ will be presented for the extreme $d_{p}$ and $G_{T_m}$ values, respectively. We consider end-to-end WPT systems, hence, when we focus on the sample times at which minima and maxima of $P_{H_p}$ and $P_{H_m}$ occur, the very same instant must be considered for $P_{T}$ due to convenience. For instance, minimum $G_{T_m}$ occurs at 00:31:50, therefore, time-varying $P_{T}$ at that very second must be considered when generating random $P_{T}$ = $P_{E_l}$ with the Monte Carlo method.

\begin{figure*}[h]
\centering
\includegraphics[width=0.52\columnwidth]{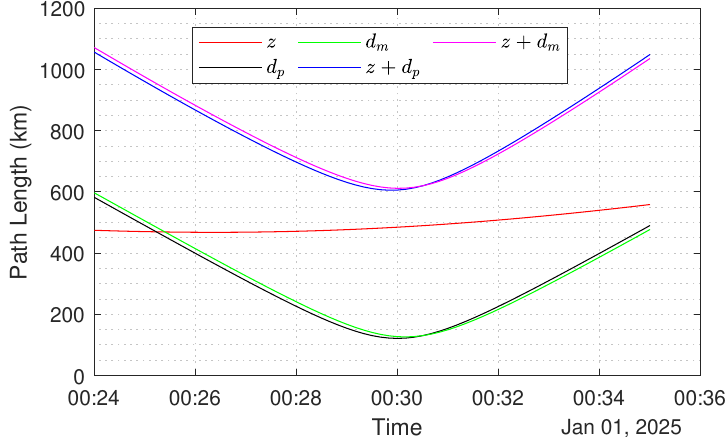}
\caption{Time-varying point-to-point path lengths ($z$, $d_p$, $d_m$) and end-to-end path lengths ($z+d_p$, $z+d_m$) that are used for the computation of end-to-end hybrid harvested power.}
\label{fig8}
\vspace{-0.2cm}
\end{figure*}

\newpage
The point-to-point and end-to-end path lengths are exhibited in Fig. \ref{fig8}. It should be noted that the maximum end-to-end distance is 1071.7 km, which is for SPS to Malapert Mountain, and hence the corresponding path delay can be computed as 3.6 ms since it is assumed that the harvested power is directly used as a relay power without using any battery system. Thus, we neglect the impact of the path length-related delays as our sampling time is 10 seconds.

The maximum and minimum $d_{p}$ values are computed as 597.0 km at 00:24:00 and 121.34 km at 00:30:00, respectively. On the other hand, the maximum and minimum $G_{T_m}$ are 38.37 dB at 00:24:00 and -39.92 dB at 00:31:50, respectively. 

\begin{figure*}[h]
\centering
\subfloat[]{
	\label{subfig:9a}
	\includegraphics[clip, scale=0.68]{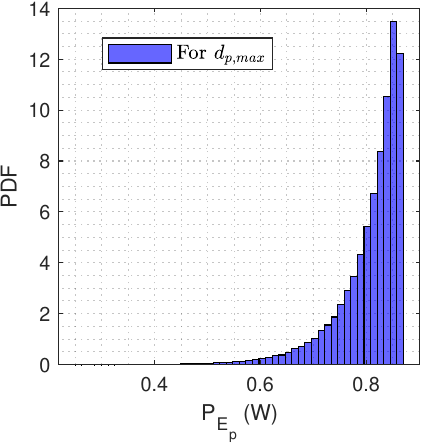}
	 }
\subfloat[]{
	\label{subfig:9b}
	\includegraphics[clip, scale=0.68]{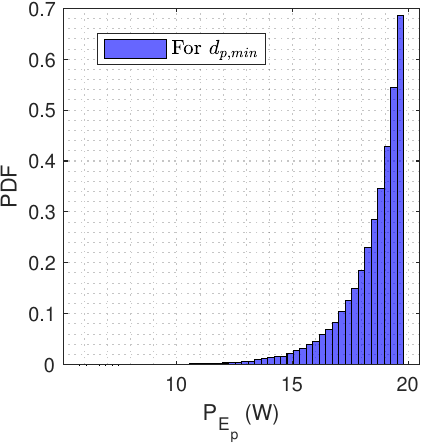}
	 }
\\
\subfloat[]{
	\label{subfig:9c}
	\includegraphics[clip, scale=0.68]{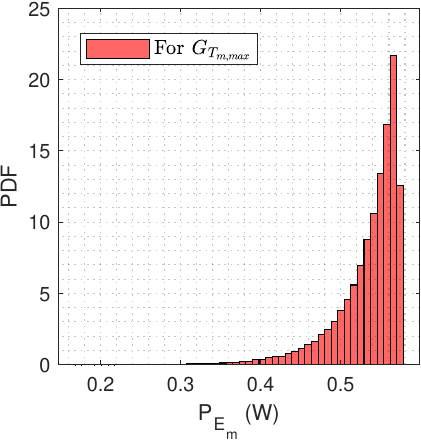}
	 }
\subfloat[]{
	\label{subfig:9d}
	\includegraphics[clip, scale=0.68]{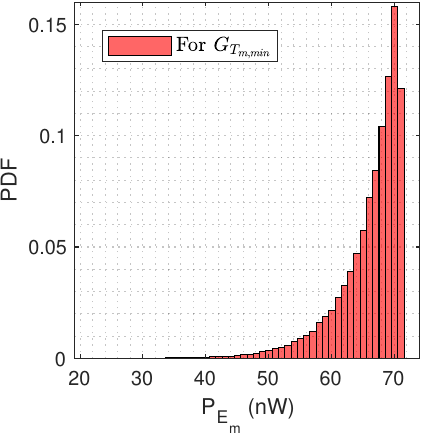}
	 }
\caption{Stochastic behaviours of (a) harvested power $P_{E_p}$ for the maximum distance $d_{p}$, (b) harvested power $P_{E_p}$ for the minimum distance $d_{p}$, (c) harvested power $P_{E_m}$ for the maximum gain $G_{T_{m}}$, and (d) harvested power $P_{E_m}$ for the minimum gain $G_{T_{m}}$.}
\label{fig9}
\vspace{-0.15cm}
\end{figure*}

The statistical distributions of $P_{E_p}$ and $P_{E_m}$ for the extreme values of $d_{p}$ and $G_{T_m}$, respectively, are presented in Fig.\ref{fig9}. The mean values of $P_{E_p}$ are 807.1 mW and 18.41 W for $d_{p,max}$ and $d_{p,min}$, respectively. On the other hand, the mean values of $P_{E_m}$ are 534.3 mW and 66.06 nW for $G_{T_{m,max}}$ and $G_{T_{m,min}}$, respectively.

\section{Conclusion}
In our hybrid WPT system model, end-to-end analyses are conducted. Firstly, the adverse effects of the misalignment fading are ignored and the time-varying harvested powers of the laser-based WPT are computed in the first hop. The harvested power is maximized when the distance between SPS and LLO satellite is minimized, and it is calculated as 331.94 kW, however, when the random pointing error is considered, the mean of the harvested power reduces to 309.49 kW for the same distance.

In the final hop, the harvested power extracted by the solar array attached to the LLO satellite is consumed entirely as relay power. Two lunar regions in which two identical parabolic antennas are considered, however, the RF-based WPT system between the LLO satellite and LSP utilizes a full-tracking module and the other system, in which a dish antenna located at Malapert Mons region uses a half-tracking module. In the perfectly aligned end-to-end hybrid WPT system, 19.80 W and 573.7 mW of maximum harvested powers are evaluated at LSP and Mountain Malapert, respectively. When the misalignment fading in the end-to-end system is considered, the mean of the maximum harvested powers degrades to 18.41 W and 534.4 mW for the former and latter hybrid WPT links. 

\section*{Acknowledgment}
This work was supported in part by the Tier-1 Canada Research Chair program.

\bibliographystyle{IEEEtran}
\bibliography{References}


\end{document}